\documentclass[12pt]{article}
\usepackage{graphicx}
\usepackage{fullpage}
\usepackage{bm}
\usepackage{color, cite}
\usepackage{amsmath, amssymb, amsfonts}

\newcommand{\ph}{|\phi(s)\rangle}
\newcommand{\ps}{|\psi(s)\rangle}
\newcommand{\xb}{\langle x}

\newcommand{\xpk}{x' \rangle}
\newcommand{\dx}{\partial_{\mathbf x}}
\newcommand{\dxs}{\partial_{x}}
\newcommand{\ex}{\mbox{exp}}
\newcommand{\eif}{\mbox{if }}

\begin{document}

\title{The effect of exclusion on nonlinear reaction diffusion system in inhomogeneous media}
\author{{Trilochan Bagarti$^1$},{~Anupam Roy$^2$\footnote{Microelectronics Research Center, J J Pickle Research Campus, The University of Texas at Austin, Texas 78758, USA}},{~K. Kundu$^1$\footnote{E-mail: kundu@iopb.res.in}}, {~and B. N. Dev$^2$}, \\$^1$Institute of Physics,\\Sachivalaya Marg, Bhubaneswar-751005, India. \\$^2$Department of Materials Science,\\Indian Association for the Cultivation of Science,\\2A and 2B Raja S. C. Mullick Road,\\Jadavpur, Kolkata 700032, India.}

\maketitle

\begin{abstract}
   We study a minimal model to understand the formation of clusters on surfaces in the presence of surface defects. We consider reaction diffusion model in which atoms undergoes reactions at the defect centers to form clusters. Volume exclusion between particles is introduced through a drift term in the reaction diffusion equation that arises due the repulsive force field produced by the clustering atoms. The volume exclusion terms can be derived from master equation with a concentration dependent hopping rate. Perturbative analysis is performed for both cross-exclusion and self-exclusion one dimensional system. For two dimension numerical analysis is performed. We have found that the clusterization process slows down due to exclusion. As a result the size of the clusters reduces. In this model reaction scheme has algebraic nonlinearity and plausible mechanism is also given.
\end{abstract}

\section{Introduction}
Cluster formation at nanoscales induced by surface defects has been studied extensively in recent times (\cite{sgar,omi,ogi,kim,xie,kim2,das,sch,sch2,trilo1} and the references therein). It has been found that step edges\cite{sgar,omi,ogi,sekar}, dislocations\cite{kim,xie,kim2,das} and domain boundaries\cite{sch,sch2} play a very crucial role in cluster formation. In our recent paper it is shown that when Ge is deposited on Si surfaces preferential growth occurs at surface defects and domain boundaries \cite{trilo1}. A reaction diffusion model is proposed which qualitatively explains the cluster formation \cite{trilo2}. In this model surface defects and domain boundaries are taken as localized reaction centers in the form of point defects and ring defects. Furthermore, simple linear form of the reaction is considered. This can be justified under certain approximation. In the studied model clusters and adatoms were allowed to diffuse normally with intrinsic diffusion coefficients \cite{trilo1,trilo2}.

Reaction diffusion models in presence of defects has been studied earlier to model a number of phenomena such as trapping of exciton in a crystal at a defect, recombination of electron and hole and soliton and antisoliton \cite{hav}. In these works trapping reactions have  been studied in which the reactants get absorbed completely or partially at trapping sites (i.e. reaction centers). The authors have focused primarily on the statistical properties such as long time behavior and self segregation \cite{tait, jay, gras, thm, guil}. Furthermore, these models describe a non-interacting system of particle undergoing reactions in a disordered media. Here we would like to emphasize that when we say an ``interacting`` reaction diffusion system, we simply mean interaction between particles other than the reactive interactions (or simply reactions). 

Our main aim in this article is two fold. In the first place, we plan to study the effect of exclusion in the formation of cluster induced by surface defects. The effect of exclusion in a multispecies reaction-diffusion system in the presence of disorder has not been studied so widely. This exclusion effect arises due to repulsion among diffusing particles. For the same kind of particles, this repulsive effect is incorporated in a mean  field way in their diffusion coefficients, which is an experimentally determined quantity \cite{chaklub}. But, there is also repulsion among dissimilar species. So, this must be taken into account at least in a mean field way in any reaction-diffusion system.

Exclusion effect in homogeneous reaction diffusion systems has been studied by a number of authors. In lattice models exclusion is incorporated by restricting the occupancy of a site strictly to a single particle. Recently, it is shown that in a lattice system with contact interactions there could be discrepancies between the lattice and their corresponding continuum model. This arises because in the continuum model the diffusion constant becomes dependent on the concentration which may take unphysical values for different lattice types and the chosen interaction neighborhood\cite{frndo}. However, the continuum model agrees well for mild to moderate contact interaction strength. 

If we look at normal diffusion the diffusivity is independent of the concentration of the diffusing particles. However, in a multiparticle system in which there is interaction between the diffusing particles the diffusivity can depend on the concentration. In such systems anomalous diffusion might be observed. It was shown that the critical behaviors of non-equilibrium absorbing phase transitions under particle conservation are affected when excluded volume interaction is incorporated \cite{kwon}.  Experimental observations have established that all concentration dependent diffusion process leads to anomalous diffusion \cite{mkuntz1, mkuntz2}. In an interacting multiparticle system, concentration dependent diffusion coefficient appears naturally \cite{tdfrank1}. Nonlinear Fokker-Planck equation has been studied in the past that describes the stochastic motion of a particle in a media whose drift and diffusion terms depends on the probability density of the particle\cite{tdfrank1, tdfrank2, lisa1,lisa2,mtz}. In the model considered here we have two coupled Smoluchosky equations \cite{vanK} in which repulsive force on any one type of particles is generated by other species. In the developed Smoluchosky equations the repulsive force on a given type of particles is assumed to be generated by the concentration gradients of the other species. For a single species system, equation studied here is same in form , developed by other authors \cite{lisa1, lisa2}. 

On the same note repulsive interaction between particles can also be seen as an exclusion effect as the repulsive force originates from an effective field produced by other particles on a tagged particle. Note that any two particle cannot occupy the same position at the same time. Hence, this effect can be introduced by a repulsive interaction between the particles (i.e. hard core repulsion) as it is done here. We would also like to emphasize here that this is an alternative way of incorporating exclusion effects in mean field equations. In this article we study a reaction diffusion system in the presence of exclusion and disorder. This type of approach has been taken to understand chemotaxis in biological problems\cite{niraj1, niraj2}.  Furthermore, exclusion processes on lattice has been studied extensively in the past to model problems in physics, chemistry and biology. It is also shown how these reaction diffusion equation can be derived using microscopic principles from the master equation \cite{vanK, radek}. 

Another important feature here is that the incorporation of nonlinear cluster formation reaction scheme. Since there is no proven reaction scheme for the formation of nanoclusters on Si surfaces, we use algebraic nonlinearity in the reaction scheme. The relevant chemical kinetic equations are derived in Appendix (\ref{sec:app2}).

The organization of the paper is as follows. In section \ref{sec:2} we discuss the theoretical model. A perturbative analysis of one-dimensional system is also presented. In Section \ref{sec:3} we study the effect of exclusion in a simple diffusion process in the presence of a trap site at the origin. We show here that self-exclusion gives rise to concentration dependent diffusion coefficient. We draw important conclusions about the formation of clusters in the presence of exclusion from this simple set up. Numerical results for both one dimensional and the original model of two dimensions are discussed in Section \ref{sec:4}. Conclusions are presented in Section \ref{sec:5}. Further scope of this work is also discussed there.
\section{Theoretical model}
\label{sec:2}
We consider a reaction diffusion process on a flat surface on which reaction occurs only in the neighborhood of reaction centers. Reaction centers are localized regions on the surface where we allow the reaction to take place. Away from the reaction centers there is only diffusion. At a reaction center we assume $\eta$ adatom combine to form a cluster. The coupled reaction diffusion equation is given by
\begin{eqnarray} 
\partial_t S(\mathbf{x},t) &=& \dx (D_s \dx S(\mathbf{x},t)+\epsilon S(\mathbf{x},t) \dx P(\mathbf{x},t)) \nonumber \\
&& - K_f(\mathbf{x}) S(\mathbf{x},t)^\eta + K_b(\mathbf{x}) P(\mathbf{x},t)+ J(\mathbf{x},t),
\label{eq:1a}
\end{eqnarray}
\begin{eqnarray}
\partial_t P(\mathbf{x},t) &=& \dx (D_p \dx P(\mathbf{x},t)+\epsilon P(\mathbf{x},t) \dx S(\mathbf{x},t)) \nonumber \\
&& - K_b(\mathbf{x}) P(\mathbf{x},t) + K_f(\mathbf{x}) S(\mathbf{x},t)^\eta.
\label{eq:1b}
\end{eqnarray}
We assume that there is no external flux, $J(\mathbf{x},t)=0$ and the initial conditions are given by $S(\mathbf{x},0)=1$ and $P(\mathbf{x},0)=0$. These equation are supplemented by appropriate boundary conditions. The diffusion and the drift terms in  Eq. (\ref{eq:1a} )and Eq. (\ref{eq:1b}) can be derived from the master equation (see Appendix \ref{sec:app1}). For $\epsilon=0$ the process is purely diffusive and describes a noninteracting system of particles. When $\epsilon \ne 0$ it describes a system in which particles of different species interact via volume exclusion. We have modeled this through an additional drift term for each species that depends on the gradient of concentration of the other species. This can be pictured in the following manner. Consider an adatom in the vicinity of a cluster. Due to thermal noise the diffusion term can be clearly written as $D_s \dx^2 S$. It is to be noted that in Fickian diffusion arising from the nonuniformity of the chemical potential the form of the diffusive term remains same, except that self-diffusion coefficient $D_s$ is replaced by cooperative diffusion coefficient \cite{chaklub}. In addition to this the adatom experiences a repelling force $(\epsilon>0)$ $ -\epsilon \dx P$ which appears as an additional drift term in  Eq. (\ref{eq:1a}) and similarly in Eq. (\ref{eq:1b}) . We note here that the surface defects help reaction to occur forming clusters in its neighborhood. On the other hand the cluster repels adatoms, therby preventing them to reach the defect site. So, clearly the formation of cluster involves a competition between these two counter processes. The exclusion of one species of particle by the other species of particles  we call here as cross-exclusion. When exclusion of a particle by their own kind is involved we will call it self-exclusion. In our model we have not included the self-exclusion-terms as it will only make the diffusion coefficient dependent on concentrations. Later in section \ref{sec:3} we will consider a trapping reaction at a static defect to examine the effect of self-exclusion.

Our next aim is to analyze the solution of these coupled equations for a perturbative exclusion effect with keeping the reaction scheme linear as it is done in our earlier work\cite{trilo2}. We further consider a single point defect at the origin. Note that if $\epsilon=0$ and $\eta=1$ the above equation becomes linear. We can write $K_f(\mathbf{x}) = k_f \delta(\mathbf{x})$ and $K_b(\mathbf{x}) = k_b \delta(\mathbf{x})$. The above equation for the linear case with a single defect is exactly solvable and we get.
\begin{equation}
S(\mathbf{x},t) =1 - \frac{k_f}{2\sqrt{D_s} k} H_s(\mathbf{x},t),
\label{eq:2a}
\end{equation}
\begin{equation}
P(\mathbf{x},t) =\frac{k_f}{2\sqrt{D_p} k} H_p(\mathbf{x},t),
\label{eq:2b}
\end{equation}
where $k = (k_f/\sqrt{D_s}+k_b/\sqrt{D_p})/2$ and the function $H_{\alpha}(\mathbf{x},t)$ is given by
\begin{equation}
H_{\alpha}(\mathbf{x},t) = \mbox{erfc}(\frac{|\mathbf{x}|}{2\sqrt{D_{\alpha} t}})- \mbox{exp}(|\mathbf{x}| k/D_{\alpha}+k^2 t) \mbox{erfc}(\frac{|\mathbf{x}|}{2\sqrt{D_{\alpha} t}}+k t).
\label{eq:3}
\end{equation}
For the nonlinear case i.e. for finite value of $\epsilon$ an analytical solution to the above set of equations cannot be obtained in a straight forward way.

Let us consider the nonlinear case $\epsilon \ne 0$ and $\eta = 1$ in one dimension with a single defect at the origin. We want to see the effect of a small exclusion ($0<\epsilon \ll 1$) term. We assume that the solution to  Eq. (\ref{eq:1a}) and Eq. (\ref{eq:1b}) can be written as \cite{nayf}
\begin{equation}
S = S_0 + \epsilon S_1 + \ldots + \epsilon^{n-1} S_{n-1}+O(\epsilon^n),
\label{eq:4a}
\end{equation}
\begin{equation}
P = P_0 + \epsilon P_1 + \ldots + \epsilon^{n-1} P_{n-1}+O(\epsilon^n),
\label{eq:4b}
\end{equation}
where $S_0$ and $P_0$ are solutions of  Eq. (\ref{eq:1a}) and Eq. (\ref{eq:1b}) with $\epsilon=0$ and is given by  Eq. (\ref{eq:2a}) and Eq. (\ref{eq:2b}) in one dimension. We can expand the  Eq. (\ref{eq:1a}) and Eq. (\ref{eq:1b}) in a regular perturbation series in powers of $\epsilon$, the resulting equations of order $n$ will be the reaction diffusion equation with $\epsilon = 0$ and a source (sink) terms centered at the defect sites that are functions of solution of order $(n-1)$ equations. The general $n$ order equation can be written as
\begin{subequations}
\begin{equation}
\partial_t S_n(x,t) = D_s \dxs^2 S_n(x,t) - K_f(x) S_n(x,t) + K_b(x) P_n(x,t)+ J_{s,n}(x,t),
\label{eq:ordn1}
\end{equation}
\begin{equation}
\partial_t P_n(x,t) = D_p\dxs^2 P_n(x,t) - K_b(x) P_n(x,t) + K_f(x) S_n(x,t)+J_{p,n}(x,t),
\label{eq:ordn2}
\end{equation}
\end{subequations}
where $J_{s,0}=0$ and $J_{p,0}=0$, $J_{s,n}(\mathbf{x},t)$ and $J_{s,n}(\mathbf{x},t)$ are source (sink) functions that are written in terms of $S_{n-k}, P_{n-k},\dx S_{n-k}$ and $\dx P_{n-k}$ for $1<k<n$. Although these equations are linear, solving order by order is still very difficult due to the complicated source (sink) terms on the right hand side. It is also not our aim to find a perturbative solution of the problem at this point. We can gain ample insight by replacing  Eq. (\ref{eq:1a}) and Eq. (\ref{eq:1b}) by a simpler set of equations. 

For small $\epsilon$ we can make the following approximation in  Eq. (\ref{eq:1a}) and Eq. (\ref{eq:1b})
\begin{equation}
\epsilon \dx S \simeq \epsilon \dx S_0,~~\epsilon \dx P \simeq \epsilon \dx P_0.
\label{eq:5}
\end{equation}
The resulting equations are a set of linear equations with variable coefficients. However, theses equations are still far from being solvable. To simplify it further we shall use the properties of the functions $S_0(x,t)$ and $P_0(x,t)$. The gradients, $\dxs S_0(x,t)$ and $\dxs P_0(x,t)$ are odd functions in $x$ and have a finite discontinuity at the reaction center $x=0$. So, we have $\dxs S_0(0^-,t) = -\dxs S_0(0^+,t)$ and $\dxs P_0(0^-,t)=-\dxs P_0(0^+,t)$. Also we have $|\dxs S_0(x,t)|$ and $|\dxs P_0(x,t)|$ monotonically decreasing for $x \in (0,-\infty)$ or $x \in (0, \infty)$ and as $x\rightarrow \pm \infty$, $\dxs S_0(x,t)=\dxs P_0(x,t)=0$. Therefore there exist  $x=x^*, y^*>0$ such that $0 \leq |\dxs S_0(x^*,t)| \leq |\dxs S_0(0^+,t)|$ and $0 \leq |\dxs P_0(y^*,t)| \leq |\dxs P_0(0^+,t)|$ for all $t>0$.

\begin{figure}
\begin{center}
\includegraphics[width=10cm]{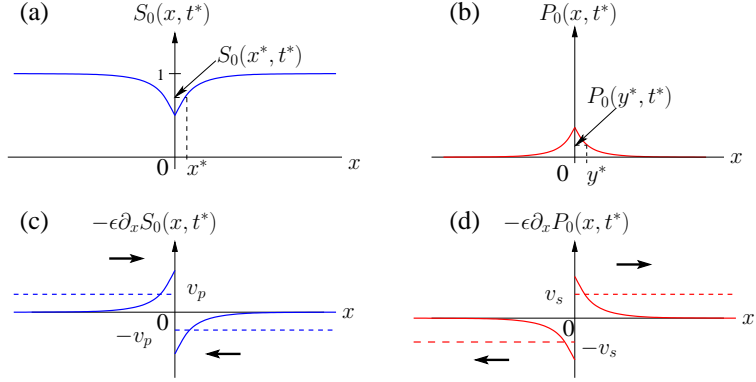}\\
\caption{Schematic diagram: (a) adatom concentration $S_0(x,t^*)$, (b) cluster concentration $P_0(x,t^*)$, (c) exclusion term $-\epsilon \dxs S_0(x,t^*)$ and (d) exclusion term $-\epsilon \dxs P_0(x,t^*)$. Thick horizontal arrows in (c) denotes the direction of the repulsive force on the cluster particles P due to the adatoms. In (d) it denotes the direction of repulsive force on the adatoms S due to the clusters. $v_s = |\epsilon \dxs P_0(y^*,t^*)|$ and $v_p = |\epsilon \dxs S_0(x^*,t^*)|$ are drift speed of adatoms and cluster arising due to exclusion . Time $t^* \in [0,t]$ and $x=x^*>0 \mbox{ and } y^*>0$ are points where the concentrations have the  mean value.}
\label{fig:schematic}
\end{center}
\end{figure}

We can make further approximations so that the gradients in  Eq. (\ref{eq:5}) can be replaced by constants which is valid in some time interval $[0,t]$. 
Let us choose $x^*, y^*>0$ and $t^* \in [0,t]$ such that it satisfies $H_s(x^*,t^*) = H_s(0,t^*)/2 $ and $H_p(y^*,t^*) = H_p(0,t^*)/2 $ (see Fig. \ref{fig:schematic}). 
Solution of these equations give the values of $x^*$and $y^*$ at which the concentrations has its mean value at some time $t^*$. In our calculations we have taken $t^*=t/2$. Equation Eq. (\ref{eq:5}) can now be written as
\begin{equation}
\epsilon \dx S \simeq v_p \mbox{sgn}(x), \epsilon \dx P \simeq -v_s \mbox{sgn}(x),
\label{eq:6}
\end{equation}
where $v_s = |\epsilon \dx P_0(y^*,t^*)|$ and $v_p = |\epsilon \dx S_0(x^*,t^*)|$ and $\mbox{sgn}(x) = 2(\theta(x)-1/2)$  accounts for the discontinuity at the reaction center $x=0$ ($\theta(x)$ is the Heaviside step function) . Since we are replacing monotonically decreasing functions by constants the approximation Eq. (\ref{eq:6}) is valid in the neighborhood of the reaction center for a small time interval $[0,t]$. Note that in this approximation all the particles are moving into or away from the reaction center at constant speeds $v_s$ and $v_p$ but in the actual case this is not true when the gradients are monotonically decreasing. However, this overestimation of $v_s$ and $v_p$ will not alter the basic physics of the problem. We obtain the following simplified reaction diffusion equations
\begin{eqnarray} 
\partial_t S(x,t) &=& D_s \dxs^2 S(x,t)-v_s \mbox{sgn}(x)\dxs S(x,t) \nonumber\\
&& - (k_f+2v_s)\delta(x) S(x,t) + k_b\delta(x) P(x,t),
\label{eq:7a}
\end{eqnarray}
\begin{eqnarray}
\partial_t P(x,t) &=& D_p \dxs^2 P(x,t)+v_p \mbox{sgn}(x) \dxs P(x,t) \nonumber \\
&& - (k_b-2v_p)\delta(x) P(x,t) + k_f\delta(x) S(x,t).
\label{eq:7b}
\end{eqnarray}
Here we note that, two very interesting features arise due to the effect of exclusion. First, it gives an extra drift term with velocity which is either into or away from the defect site. Secondly, it modifies the reaction rates and the reaction terms become different for two reacting species breaking constraints of our kinetic scheme (see  Eq. (\ref{eq:1a}) and Eq. (\ref{eq:1b})). Let us define $\tilde{k}_f =k_f+2v_s$ and $\tilde{k}_b =k_b-2v_p$. The  Eq. (\ref{eq:7a}) and Eq. (\ref{eq:7b}) after Laplace transform can be written in an abstract notation \cite{eco} as the following
\begin{equation}
\ph = G_s(s)\left[ -\tilde{K}_f\ph + K_b \ps +  |J(s)\rangle\right], 
\label{eq:8a}
\end{equation}
\begin{equation}
\ps = G_p(s)\left[ -\tilde{K}_b\ps + K_f \ph \right],
\label{eq:8b}
\end{equation}
where $\xb \ph = \phi(x,s)$ and $\xb \ps = \psi(x,s)$ are the Laplace transform of $S(x,t)$ and $P(x,t)$ respectively. The flux term appears due to the initial condition $S(x,0) = 1 =\xb|J(s)\rangle$. The reaction operators are defined by $\delta(x-x') k_\alpha \delta(x) = \xb|K_\alpha|\xpk$ and $\delta(x-x') \tilde{k}_\alpha\delta(x) = \xb|\tilde{K}_\alpha|\xpk$. The Green's functions $G_s(s)$ and $G_p(s)$ are defined by
\begin{equation}
G_s(s) =\left[s-D_s\dx^2+v_s \mbox{sgn}(x) \dx\right]^{-1},
\label{eq:9a}
\end{equation}
\begin{equation}
G_p(s) =\left[s-D_p\dx^2-v_p \mbox{sgn}(x) \dx\right]^{-1}.
\label{eq:9b}
\end{equation}
The expressions for the Green's functions in  Eq. (\ref{eq:9a}) and Eq. (\ref{eq:9b}) are given in \ref{sec:app3}. Next consider diffusion of adatoms on a surface without defects. If we choose the initial concentration $S(x,0) = \delta(x)$, we know that it will evolve as Gaussian as there is no cluster formation. Now suppose that we introduce a force field at the origin such that it gives rise to a constant drift velocity $v_s$ in the outward direction. The diffusion of adatoms can be described by the following equation
\begin{equation}
\partial_t S(x,t) = \dxs \left[ D_s \dxs-v_s \mbox{sgn}(x)\dxs\right] S(x,t)
\label{eq:10}
\end{equation}
In the Laplace domain we can write
\begin{equation}
\phi(x,s) = \frac{G_s(x|0)}{1+2v_sG_s(0|0)}.
\label{eq:11a}
\end{equation}
Here $\phi(x,s)$ is the Laplace transform of $S(x,t)$. For brevity the Laplace variable $s$ is kept implicit in $G_s(x|x')$. Using the expression for the Green's function Eq. (\ref{eq:grn1} and \ref{eq:grn2}) in  Eq. (\ref{eq:11a}) and taking inverse Laplace transform, the solution of  Eq. (\ref{eq:10}) can be written as
\begin{equation} 
S(x,t)=\frac{1}{\sqrt{4 \pi D_s t}}\ex\left( \frac{-(|x|-v_s t)^2}{4D_s t}\right)-\frac{v_s}{4D_s}\mbox{e}^{v_s|x|/D_s} \mbox{erfc}\left(\frac{|x|+v_s t}{2\sqrt{D_s t}} \right).
\label{eq:11b}
\end{equation}
Looking at  Eq. (\ref{eq:11b}) we note from the first term that adatoms are pushed away from the origin to a distance $v_st$. The second term has a minimum at the origin and reduces the concentration by small amount which is of the order $v_s$. 

Now returning to our reaction diffusion problem we should expect that, in the presence of exclusion, adatoms will experience an extra repulsive force which is directed outward from the center of the surface defect. From  Eq. (\ref{eq:8a}) andEq. (\ref{eq:8b}) we obtain 
\begin{equation}
\phi(x,s) = Q(x)-\frac{\tilde{k}_fG_s(x|0)Q(0)}{\Delta}-\frac{(\tilde{k}_f\tilde{k}_b-k_fk_b)G_s(x|0)G_p(0|0)Q(0)}{\Delta},
\label{eq:12a}
\end{equation}
\begin{equation}
\psi(x,s) = \frac{k_f G_p(x|0)Q(0)}{\Delta},
\label{eq:12b}
\end{equation}
where 
\begin{equation}
\Delta = 1 + \frac{\tilde{k}_f}{2D_s(-\rho_s+\gamma_s)}+ \frac{\tilde{k}_b}{2D_p(-\rho_p+\gamma_p)}+ \frac{\tilde{k}_f\tilde{k}_b-k_fk_b}{4D_sD_p(-\rho_s+\gamma_s)(-\rho_p+\gamma_p)},
\label{eq:12c} 
\end{equation}
$\rho_s = v_s/(2D_s),~\rho_p = -v_p/(2D_p),~\gamma_s=\sqrt{\rho_s^2+s/D_s}$ and $\gamma_p=\sqrt{\rho_p^2+s/D_p}$. The function $Q(x) =\xb|G_s(s)|J\rangle = 1/s$. 

\begin{figure}[!h]
\begin{center}
\includegraphics[width=6cm]{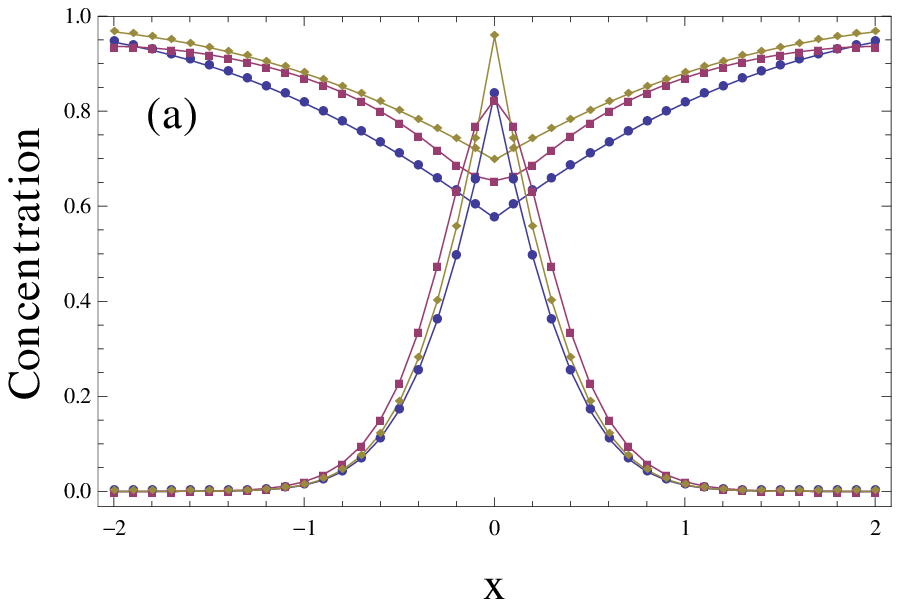}
\includegraphics[width=6cm]{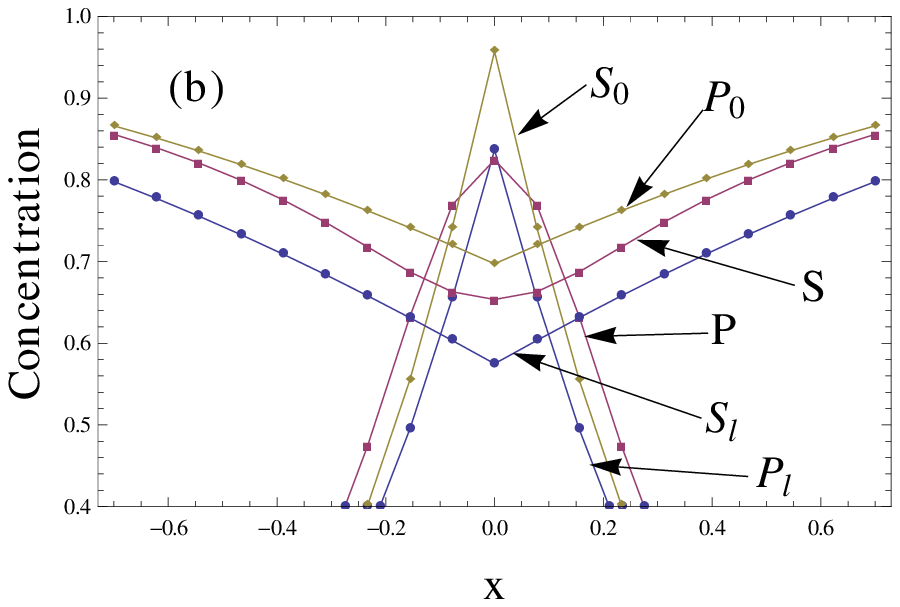}
\caption{(a) Concentrations: Numerical solution $S,P$, solution without exclusion ($\epsilon =0$) $S_0, P_0$ and solutions of the simplified  linear equation $S_l, P_l$, (b) Concentrations close to the defect at the origin.}
\label{fig:1}
\end{center}
\end{figure}

The inverse Laplace transformation of $\phi(x,s)$ and $\psi(x,s)$ is performed numerically by Talbot method\cite{tal} which we denote by $S_l(x,t)$ and $P_l(x,t)$ respectively. In Fig. \ref{fig:1} (a) and \ref{fig:1} (b) we have plotted the concentrations for the case $\epsilon = 0$ denoted by $S_0, P_0$, actual numerical solution $S, P$ of  Eq. (\ref{eq:1a}) and Eq. (\ref{eq:1b}) obtained by finite difference method and the solution of the modified linear equations $S_l, P_l$. The parameters are $\epsilon = 0.1,~D_s=1,~D_p=0.25,~k_f=1.0,~k_b=0.1$ and $t=1.0$. In Fig. \ref{fig:1}(b), we note that the solution $S_l$ and $P_l$  and the numerical solution $S$ and $P$ at a point close to the defect site have reduced as compared to the bare case, i.e. $\epsilon=0$ concentrations, $S_0,~P_0$. The current due to diffusion and the drift current are in opposite directions for both adatom and cluster which effectively reduces the number of particle at the origin. However, this is true only for small value of $0<\epsilon~\ll 1$. For higher values of $\epsilon$ the diffusion and the drift currents will be comparable and the higher order terms in $\epsilon$ will also have a significant contribution (see Fig. \ref{fig:2}).
\section{Trapping reaction with self-exclusion}
\label{sec:3}
To understand the effect of self-exclusion we consider a single diffusing species with a trap at the origin. We assume that particles interact among themselves, so we introduce a self-exclusion term proportional to the density gradient as we have done in the previous section. When a particle reaches the trap site it gets trapped with a time-independent rate constant $\kappa$. The reaction diffusion equation for the problem is then
\begin{equation}
\partial_t u(x,t) = \dxs (D_0 \dxs u(x,t)+\epsilon u(x,t) \dxs u(x,t)) - \kappa \delta(x) u(x,t).
\label{eq:13}
\end{equation}
We use the initial condition $u(x,0)=1$ along with appropriate boundary conditions\cite{len}. By rearranging terms in  Eq. (\ref{eq:13}) the term due to exclusion can simply be absorbed in a concentration dependent diffusion coefficient, $D(u)=D_0(1+\epsilon u)$ (here $\epsilon$ is redefined as $\epsilon = \epsilon/D_0$). We note that with self-exclusion the current due to diffusion and drift are in the same direction as this can be can seen from the expression for the current $j(u) = -D(u)\dxs u$. For $\epsilon=0$, Eq. (\ref{eq:13}) has an exact solution \cite{guil}. Let us consider the case of a small reaction rate $\kappa \ll 1$ and $\epsilon \ne 0$. 

With no loss of generality, setting $D_0=1$ we expand  Eq. (\ref{eq:13}) in terms of perturbation series in  $\kappa$ \cite{nayf}.
\begin{equation}
u = u_0 +\kappa u_1 + \kappa^2 u_2+O(\kappa^3).
\label{eq:14}
\end{equation}
The solution $u_0=1$ satisfy the zeroth order equation. The equations for $u_1$ and $u_2$ are given by
\begin{equation}
\partial_t u_1 = (1+\epsilon) \dxs^2 u_1-\delta(x), u_1(x,0) =0,
\label{eq:15}
\end{equation}
\begin{equation}
\partial_t u_2 = (1+\epsilon) \dxs^2 u_2 - \delta(x) u_1 +\epsilon \dxs(u_1 \dxs u_1), u_2(x,0) =0.
\label{eq:16}
\end{equation}

\begin{figure}[!h]
\begin{center}
\includegraphics[width=7cm]{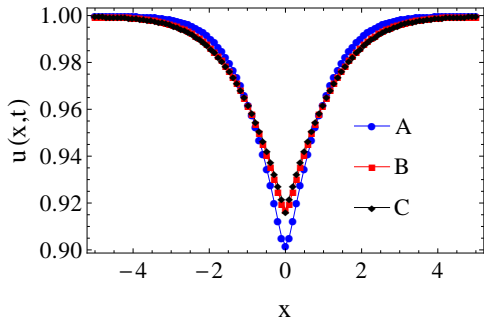}
\caption{Concentration $u(x,t)$ numerical solution (A) without exclusion $\epsilon =0.0$ and with exclusion $\epsilon=0.5$ (B) numerical solution, (C) analytical solution Eq. (\ref{eq:20}) for $\kappa = 0.2$ and time $t=1.0$. The difference between (A) and (C) at $x=0$ is $\Delta u$.}
\label{fig:6}
\end{center}
\end{figure}

Equation Eq. (\ref{eq:15}) describe diffusion with an external flux which for this case is a negative point flux at the origin. In  Eq. (\ref{eq:16}) the expression $\epsilon \dxs(u_1 \dxs u_1)$ albeit exactly known is quite complicated. It is maximum at the origin and monotonically decreases with increasing values of $|x|$ and vanishes at infinity. Inasmuch as we are interested in the behavior of the solution at a finite time, we can replace this term by a point flux at the origin without compromising the basic physics. This assumption is valid only for small $t$. We have 
\begin{equation}
\dxs(u_1 \dxs u_1) \simeq j_0 \delta(x),
\label{eq:17}
\end{equation}
where $j_0 = \lim_{x \rightarrow 0}\dxs(u_1 \dxs u_1) = (\pi+2)/(4 \pi (1+\epsilon)^2)$. Substituting  Eq. (\ref{eq:17}) in  Eq. (\ref{eq:16}) and solving  Eq. (\ref{eq:15}) and Eq. (\ref{eq:16}) we have
\begin{eqnarray}
u(x,t) &=& 1-\frac{\kappa - \epsilon \kappa^2 j_0}{2 \sqrt{1+\epsilon}}\left[ 2 \sqrt{\frac{t}{\pi}} \ex \left( \frac{-x^2}{4(1+\epsilon)t} \right) - \frac{|x|}{\sqrt{1+\epsilon}} \mbox{erfc}\left( \frac{|x|}{2\sqrt{(1+\epsilon)t}} \right)  \right] \nonumber \\
&&+\frac{\kappa^2}{4(1+\epsilon)}\left[ \left( \frac{x^2}{2(1+\epsilon)}+t\right) \mbox{erfc}\left( \frac{|x|}{2\sqrt{(1+\epsilon)t}} \right) \right. \nonumber \\
&& \left.  - |x| \sqrt{\frac{t}{\pi}} \ex \left( \frac{-x^2}{4(1+\epsilon)t} \right) \right] + O(\kappa^3).
\label{eq:20}
\end{eqnarray}
Let $\tilde{u}(x,t) = \lim_{\epsilon \rightarrow 0}u(x,t)$ be the concentration when there is no self-exclusion. Define by $\Delta u = u(x,t)-\tilde{u}(x,t)$, the difference in the concentration. From Eq. (\ref{eq:20}) we note that the concentration at the origin has increased due to the flux term $j_0$ and we have $\Delta u \simeq (1-1/\sqrt{1+\epsilon}) \kappa \sqrt{t/\pi}+\kappa^2 \epsilon j_0 \sqrt{t/(\pi(1+\epsilon))}-(1-1/(1+\epsilon))\kappa^2t/4$. For $\epsilon=0.5,~\kappa=0.2$ at $t=1.0$ we have $\Delta u=0.019$ and the corresponding numerical solutions give $\Delta u_{numerical}=0.0154$ see Fig. (\ref{fig:6}). We see that in this case the effect is exactly the opposite (compare Fig. \ref{fig:1} for the multispecies case). Furthermore, the width of the depletion zone has increased due to the increase in the diffusion coefficient by $\epsilon$.

\section{Numerical results}
\label{sec:4}
\begin{figure}
\begin{center}
\includegraphics[width=6cm]{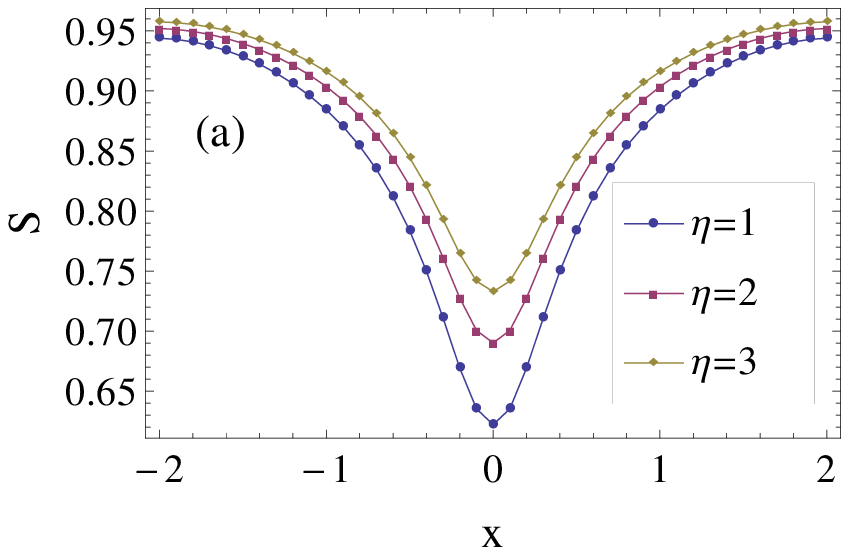}
\includegraphics[width=6cm]{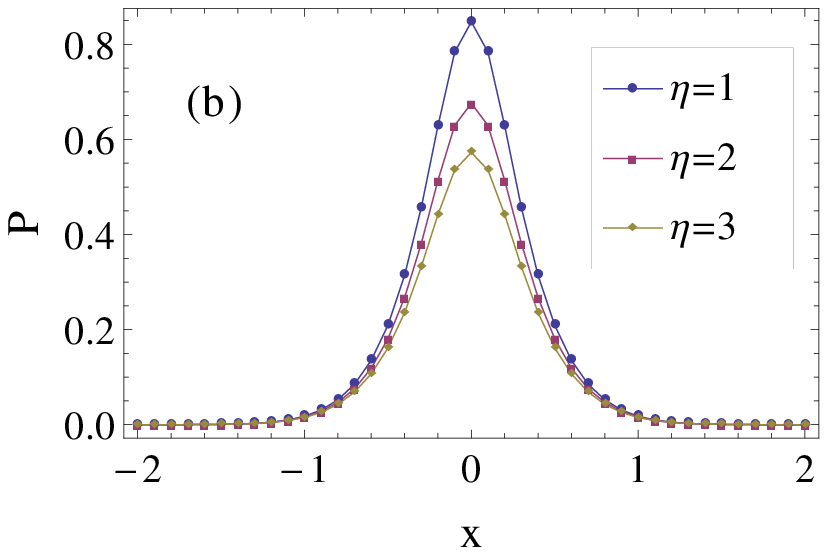}\\
\includegraphics[width=6cm]{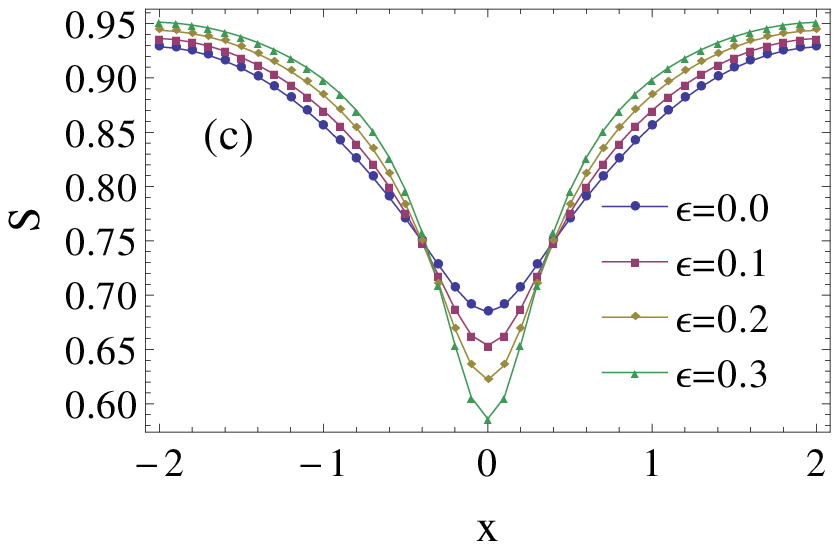}
\includegraphics[width=6cm]{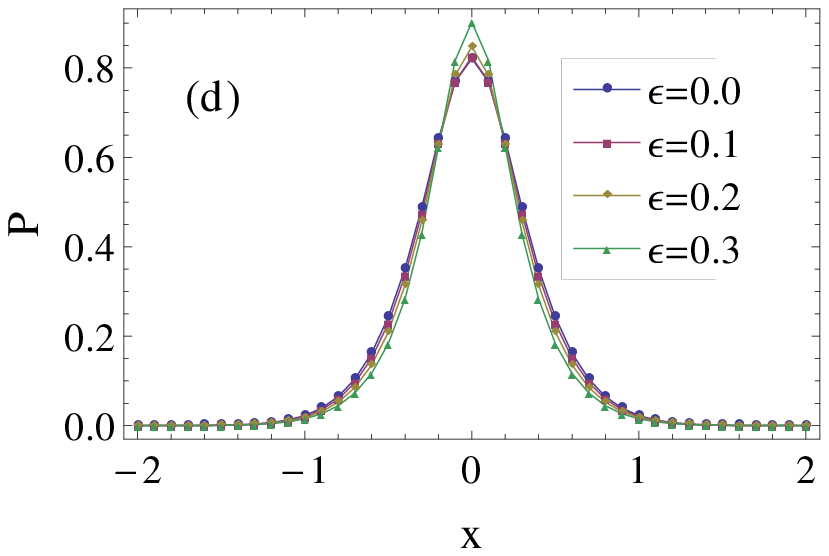}
\caption{Variation of concentration with exclusion and nonlinearity. (a) and (b) $\epsilon=0.2$ and $\eta=1,~2,~3$, (c) and (d) $\eta=1,~\epsilon=0.0,~0.1,~0.2,~0.3$. The parameters are $D_s=1.0,~D_p=0.25,~k_f=1.0$ and $ k_b=0.1$ and $t=1.0$.  }
\label{fig:2}
\end{center}
\end{figure}

\begin{figure}
\begin{center}
\includegraphics[width=8cm]{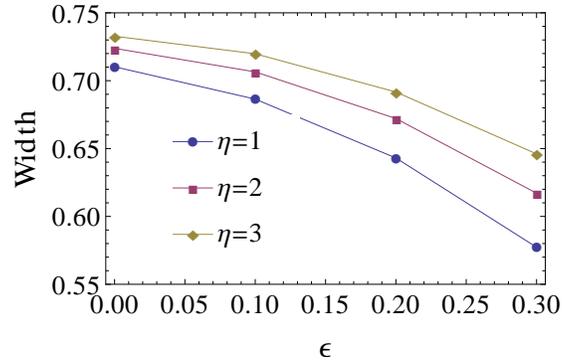}
\caption{Width of concentration $P(x,t)$ for $D_s=1.0,~D_p=0.25,~k_f=1.0$ and $ k_b=0.1$ and $t=1.0,~\eta=1,~2,~3$.}
\label{fig:3}
\end{center}
\end{figure}

\begin{figure}
\begin{center}
\includegraphics[width=6cm]{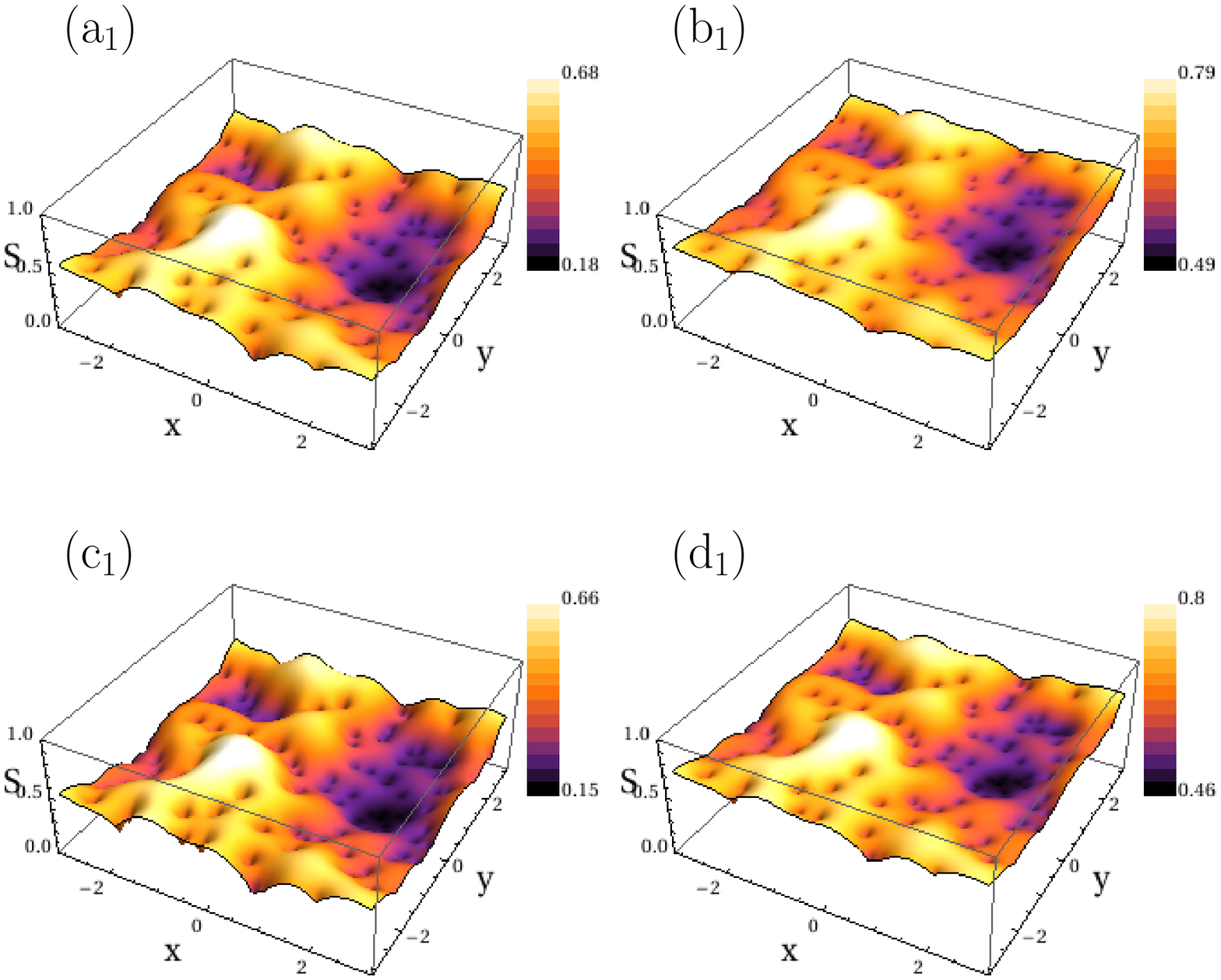}
\includegraphics[width=6cm]{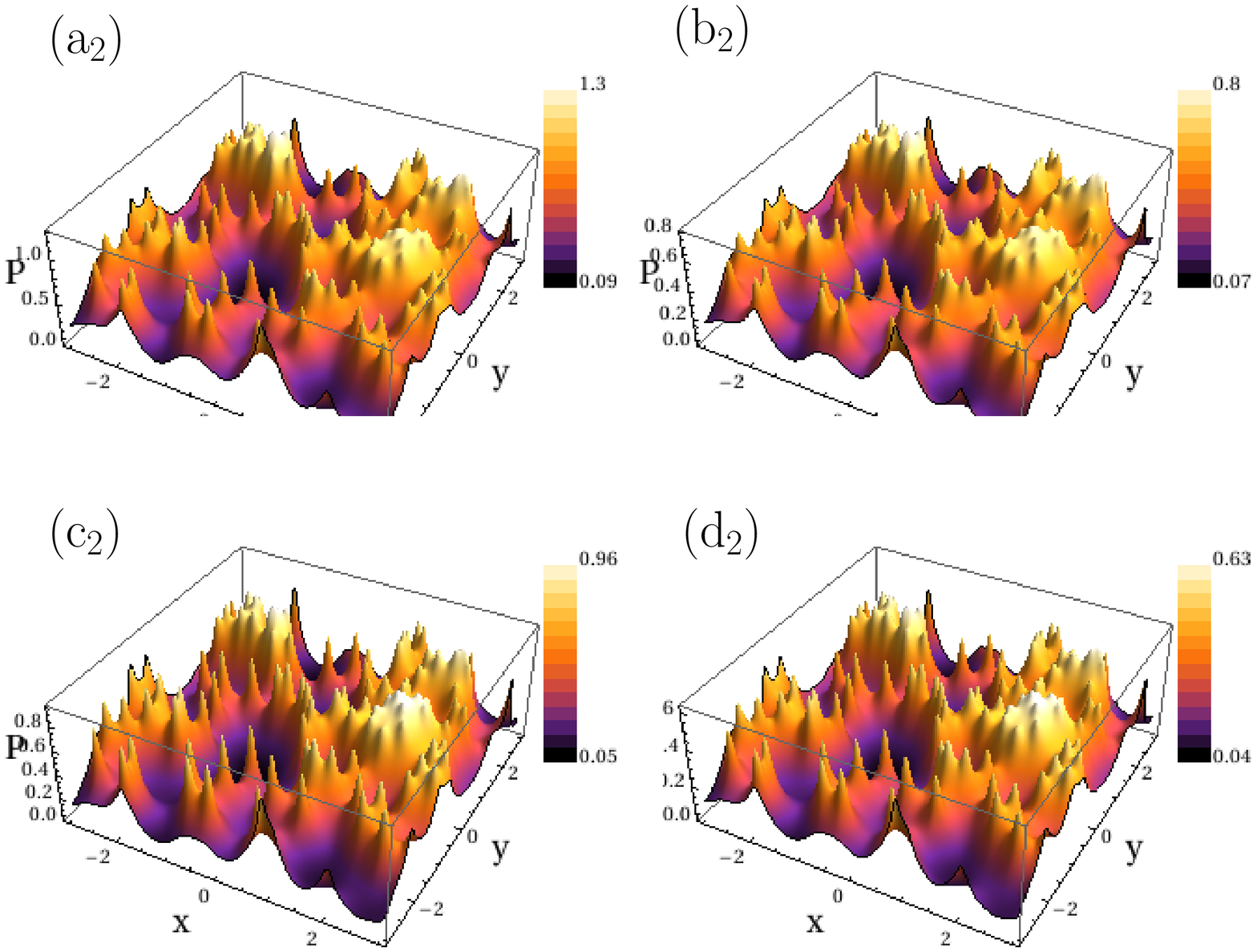}
\caption{Formation of clusters in the presence of exclusion. (a$_{1,2}$) $\epsilon=0,~\eta=1$, (b$_{1,2}$) $\epsilon=0,~\eta=3$, (c$_{1,2}$) $\epsilon=0.2,~\eta=1$, (d$_{1,2}$) $\epsilon=0.2,~\eta=3$ for $D_s=1.0,~D_p=0.25,~k_f=1.0$ and $ k_b=0.1$ and $t=1.0$.}
\label{fig:4}
\end{center}
\end{figure}

\begin{figure}
\begin{center}
\includegraphics[width=4cm]{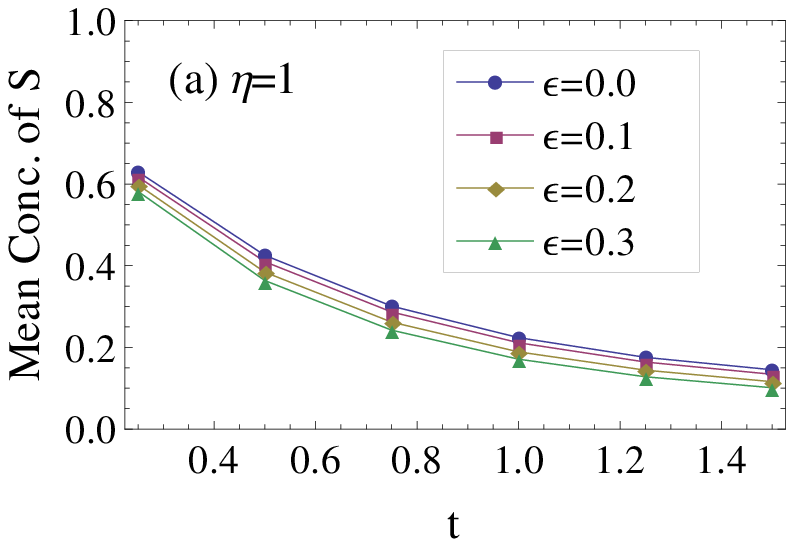}
\includegraphics[width=4cm]{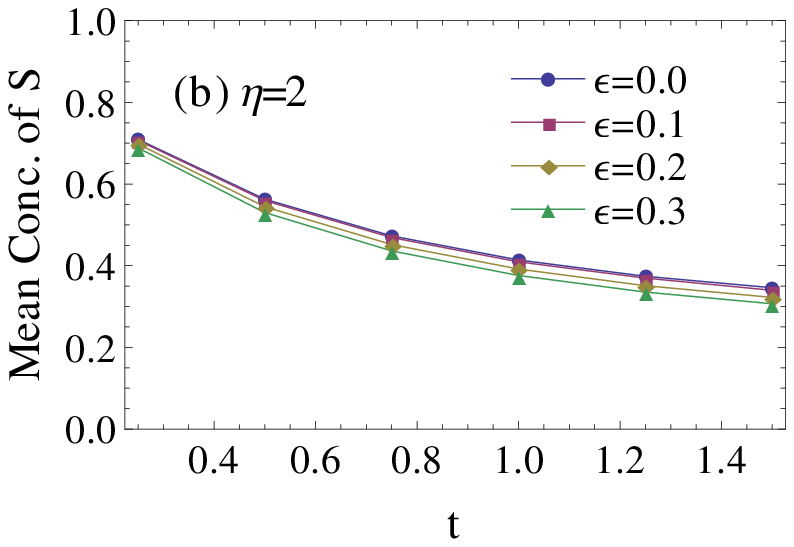}
\includegraphics[width=4cm]{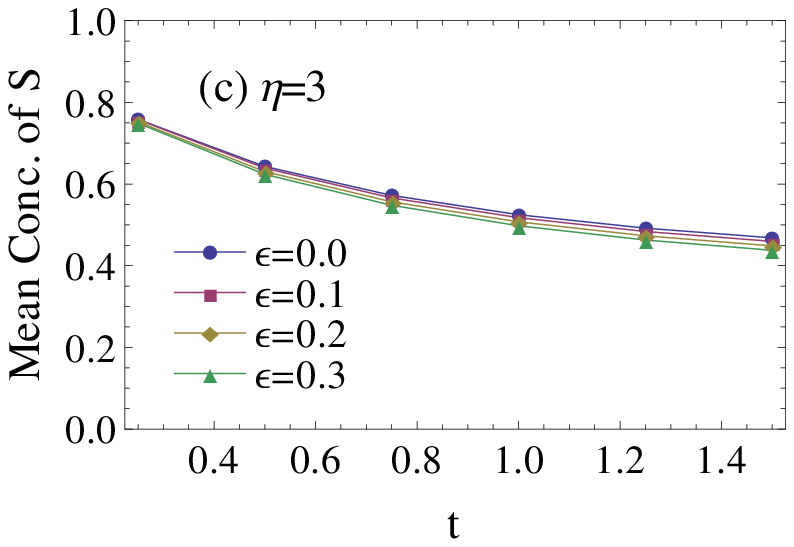}\\
\includegraphics[width=4cm]{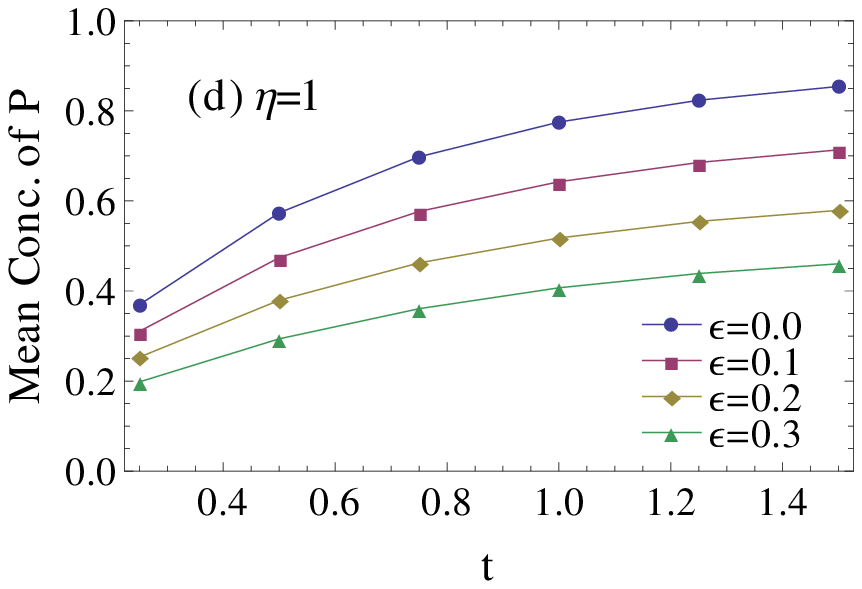}
\includegraphics[width=4cm]{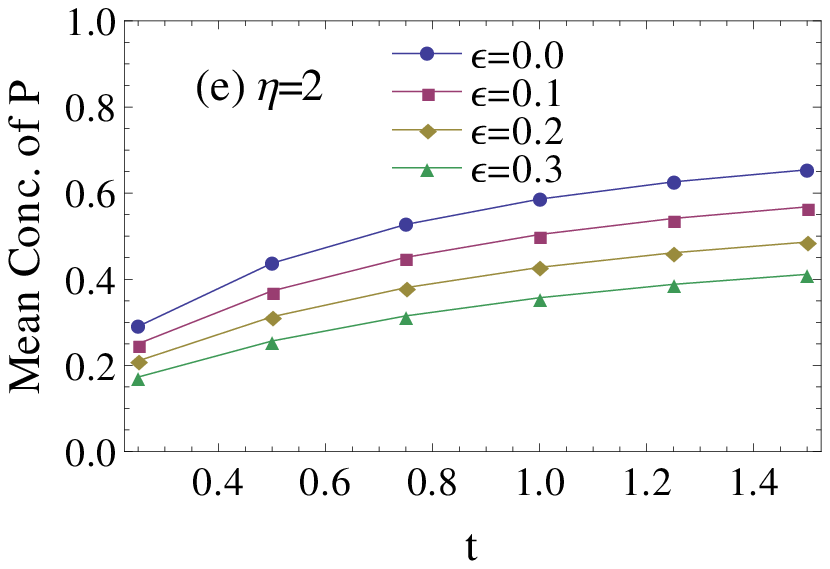}
\includegraphics[width=4cm]{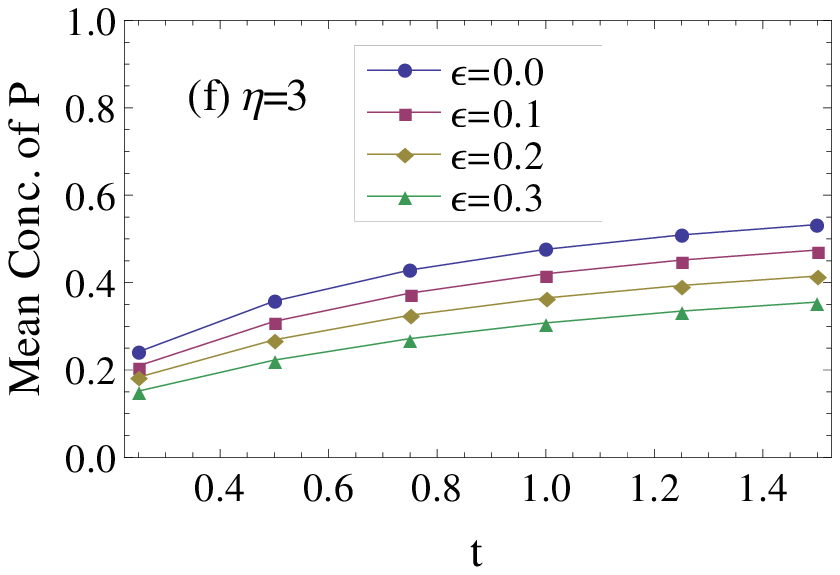}
\caption{Mean concentrations of adatoms S for (a)$\eta=1$, (b)$\eta=2$, (c)$\eta=3$ and cluster P for (d)$\eta=1$, (e)$\eta=2$, (f)$\eta=3$ and $D_s=1.0, D_p=0.25, k_f=1.0$ and $ k_b=0.1$ and $0.25 \leq t \leq 1.5$.}
\label{fig:5}
\end{center}
\end{figure}

The well known finite difference method\cite{ames} is used to to compute the solution of Eq. (\ref{eq:1a}) and Eq. (\ref{eq:1b}) numerically. In one dimension we will first examine the effect of exclusion and nonlinearity on the shape of the concentration profile with a reaction center at the origin. The parameters used are $D_s=1.0,~D_p=0.25,~k_f=1.0$ and $ k_b=0.1$. In Fig. \ref{fig:2} (a) and \ref{fig:2}(b) we have plotted concentration $S(x,t)$ for $\eta = 1,2 \mbox{ and }3$ at $\epsilon = 0.2$ and time $t=1.0$. The concentration of S is decreases with increase in $\eta$. The variation of concentrations with different values of the exclusion parameter $\epsilon = 0.0,~0.1,~0.2 \mbox{ and } 0.3$ are shown is Fig. \ref{fig:2} (c) and \ref{fig:2} (d). Here as we increase $\epsilon$, the concentration $S(x,t)$ decreases. It has already been discussed in our study of the modified linear equations (see  Eq. (\ref{eq:7a}) and Eq. (\ref{eq:7b}). It is shown that exclusion effect modifies the reaction rate at the reaction center and consequently the concentration decreases. We also note that change in concentration $P(x,t)$ with $\epsilon$ is negligible for $\epsilon \ll 1$. This is also due to the fact that change in $P(x,t)$ due to exclusion is not first order in $\epsilon$. We have also found that the width of the concentration profile of P reduces as the parameter $\epsilon$ is increased. In Fig. (\ref{fig:3}) we have calculated the FWHM for the concentration $P(x,t)$ for different values of $\eta$. This clearly indicates that exclusion or/and nonlinearity suppress the formation of clusters. 
In Fig. \ref{fig:4} we  have plotted the concentration profile of P in two dimensions. Here we have used the same set of parameters as in the one dimension case. The number of defects is $100$ which is uniformly distributed in the region $-3\leq x \leq 3$ and $-3\leq y \leq 3$. For a given randomly distributed defects we have plotted concentration $P(x,y,t)$. Here also we see that as we go from $\epsilon =0$ to $\epsilon = 0.2$ keeping $\eta =1$ fixed the concentration decreases. The concentration is also found to decrease as we increase the nonlinearity from $\eta = 1$ to $\eta = 3$.
In Fig. \ref{fig:5} we have calculated the mean concentration averaged over the randomness of defect distribution. The decay of concentration S monotonically decreases with time and the rate of its decay slows down as $\eta$ is increased. Similarly for P its mean concentration increases with time and its concentration for any given time $t$ increases with increase in $\eta$. We also note that the mean concentration for both S and P decreases with increasing $\epsilon$. This clearly suggests that both exclusion and nonlinearity suppress the formation of cluster and the effect of both is additive.
\section{Conclusion}
\label{sec:5}
This model is primarily designed to understand the formation of Ge-cluster on Si-surfaces. It is a minimal model for the said type of problems. In this work the effect of exclusion on the formation of cluster in the presence of surface defects is studied. In our reaction diffusion model we have introduced exclusion through a repulsive interaction between the particles of dissimilar nature. This repulsive force is taken to be proportional to the gradient of concentration and a set of coupled Smoluchowsky equations is obtained. In the perturbative regime a set of modified linear reaction-diffusion equation as an approximation to the actual process is considered and this gave us important understanding of the effect of exclusion on the formation of clusters in the reaction diffusion processes. The solutions to these equation are obtained using Green's function method\cite{hilde}. The most important conclusion of this work is that, both exclusion and nonlinear reaction processes considered here suppress the formation of cluster. The effect of self-exclusion on diffusion in one dimension with a trapping reaction at the origin is studied. It is found that self-exclusion can give rise to a concentration dependent diffusion coefficient as obtained in earlier works\cite{tdfrank2}. The width of the depletion zone is found to increase by $\epsilon$ (strength of exclusion). Our numerical studies in one dimension showed that the width of the cluster concentration profile decreases with increasing $\epsilon$. In two dimensions, the mean concentrations averaged over the surface disorder are calculated. It is found that the mean concentration of adatoms S  decreases monotonically where as the mean concentration of P increases monotonically in time respectively. However, for higher exclusion and nonlinearity these mean concentrations are found to decrease with the increase in exclusion and nonlinearity in the reaction scheme. 

There can be quite a few extension of the present work. For example, in this model, the repulsive potential is considered to be simple linear function of concentration. But, it is indeed possible for this potential to depend nonlinearly on concentration. This needs to be explored. Furthermore, we have considered here a simple algebraic nonlinearity in the reaction scheme. This is due to the lack of sufficient knowledge of the reaction scheme for the formation of Ge-clusters on Si surfaces. As this model can be used in other physical situations, different types of nonlinear exclusion potential and reaction scheme can be studied. These works are in progress.
\appendix
\section{Derivation of exclusion term from master equation}
\label{sec:app1}
The diffusion and the exclusion terms in  Eq. (\ref{eq:1a}) and Eq. (\ref{eq:1b}) from the master equation are derived in the following way. Let us discretize the continuous space into cells of size $\Delta x$ and denote by $n_i\geq 0$ and $m_i\geq 0$ for all $i=1,2,\ldots,N$ the number of S and P particles in the $i$-th cell. The configuration of the system can be described by the N-vectors $\mathbf{n}=(n_1,n_2,\ldots,n_N)$  and $\mathbf{m}=(m_1,m_2,\ldots,m_N)$. Let $\mathbb{N}$ be the set of natural numbers, we define the operator $H_i^{\pm}:\mathbb{N}^N\rightarrow \mathbb{N}^N$ by
\begin{eqnarray}
H_i^+(n_1,\ldots,n_{i-1},n_i,n_{i+1},\ldots,n_N) &=& (n_1,\ldots,n_{i-1},n_i-1,n_{i+1}+1,\ldots,n_N), \nonumber \\
H_i^-(n_1,\ldots,n_{i-1},n_i,n_{i+1},\ldots,n_N) &=& (n_1,\ldots,n_{i-1}+1,n_i-1,n_{i+1},\ldots,n_N), \nonumber \\
\end{eqnarray}
for all $i=1,2,\dots,N$. Note that for $i=1,N$ we use periodic boundary condition\cite{radek}. 

 We define density dependent hopping rates $W_s^{\pm}(i)$ and  $W_p^{\pm}(i)$ of a particle at the $i$th site for S and P respectively by
\begin{eqnarray}
W_s^{\pm}(i) &=& w_s(1-\mu_s (m_{i\pm1}-m_i)), \nonumber \\
W_p^{\pm}(i) &=& w_p(1-\mu_p (n_{i\pm1}-n_i)),
\end{eqnarray}
where $w_s, w_p, \mu_s$ and $\mu_p$ are constants and the superscript $\pm$ denotes hopping to the site $i\pm1$. Here we note that a particle S (say) at cell $i$ has lower hopping rate if the cell $i\pm1$ contains more number of particles P than that in the cell $i$. Similarly for P this is exactly what should be happening when there is an effect of exclusion. Let $p(\mathbf{n},\mathbf{m},t)$ be the probability of finding the system in the configuration $\mathbf{n},\mathbf{m}$, at time $t$
The master equation is given by\cite{vanK, radek}
\begin{eqnarray}
\partial_t p(\mathbf{n},\mathbf{m},t) &=&\sum_{j=1}^N(n_j+1) \left[ W_s^-(j)p(H_{j-1}^+\mathbf{n},\mathbf{m},t) +W_s^+(j)p(H_{j+1}^-\mathbf{n},\mathbf{m},t)\right]\nonumber \\
&&+(m_j+1) \left[ W_p^-(j)p(\mathbf{n},H_{j-1}^+\mathbf{m},t) +W_p^+(j)p(\mathbf{n},H_{j+1}^-\mathbf{m},t)\right] \nonumber \\
&& -\left( n_j \left[ W_s^-(j) +W_s^+(j)\right]+m_j \left[ W_p^-(j) +W_p^+(j)\right]\right) p(\mathbf{n},\mathbf{m},t) \nonumber \\ 
\label{eq:23}
\end{eqnarray}
Let $S_i = \sum_\Omega n_i p(\mathbf{n},\mathbf{m},t)$ and $P_i = \sum_\Omega m_i p(\mathbf{n},\mathbf{m},t)$ where $\sum_\Omega$ represents sum over all configurations having $n_1,\ldots,n_N,m_1,\ldots,m_N$, be the mean number of particles in the $i$th cell. Let us assume that there are no correlation between particles so that we can write $\sum_\Omega n_i n_j p(\mathbf{n},\mathbf{m},t) = S_i S_j$, $\sum_\Omega m_i m_j p(\mathbf{n},\mathbf{m},t) = P_i P_j$ and $\sum_\Omega n_i m_j p(\mathbf{n},\mathbf{m},t) = S_i P_j$. Multiplying $n_i$ through  Eq. (\ref{eq:23}) and summing over all configurations we obtain 
\begin{eqnarray}
\partial_t S_i &=& w_s(S_{i+1}+S_{i-1}-2S_i)- w_s\mu_s \left[S_{i+1}(P_i-P_{i+1})+S_{i-1}(P_i-P_{i-1})\right. \nonumber \\
&&\left.-S_i(P_{i+1}+P_{i-1}-2P_i)\right]
\label{eq:24}
\end{eqnarray}
Now we can substitute $S_i = S(x,t), P_i = P(x,t)$ and expand $S_{i\pm1}$ and $P_{i\pm1}$ in Taylor series 
\begin{eqnarray}
S_{i\pm1} \simeq S(x,t) \pm \Delta x \dxs S(x,t)+\Delta x^2 \dxs^2 S(x,t), \nonumber \\
P_{i\pm1} \simeq P(x,t) \pm \Delta x \dxs P(x,t)+\Delta x^2 \dxs^2 P(x,t),
\end{eqnarray}
and ignoring $O(\Delta x^3)$ and higher order terms, we obtain from  Eq. (\ref{eq:24})
\begin{eqnarray}
\partial_t S(x,t) &=& w_s\Delta x^2\dxs^2S(x,t)-w_s\mu_s \Delta x^2 \left[-\left( 2\dxs S(x,t) \dxs P(x,t)\right. \right. \nonumber \\
&&\left.\left.+S(x,t) \dxs^2 P(x,t) \right) -S(x,t)\dxs^2 P(x,t)\right]
\label{eq:26}
\end{eqnarray}
Now rearranging terms in  Eq. (\ref{eq:26}) we obtain
\begin{equation}
\partial_t S(x,t) = D_s \dxs^2 S(x,t) + \epsilon_s(\dxs S(x,t)\dxs P(x,t) + S(x,t) \dxs^2 P(x,t)),
\label{eq:diffEx}
\end{equation}
where $D_s = w_s \Delta x^2$ and $\epsilon_s = 2 w_s \mu_s \Delta x^2$. Similarly multiplying $m_i$ through  Eq. (\ref{eq:23}) and summing over all configurations we obtain the equation for $P(x,t)$. A point to note here is that $0<\mu_{s,p} \ll 1$, and for simplicity we have taken $\epsilon_s = \epsilon_p = \epsilon$. Using expression Eq. (\ref{eq:diffEx}), the general expression in arbitrary dimension can be written as
\begin{equation}
\partial_t S(\mathbf{x},t) = \dx\left[D_s \dx S(\mathbf{x},t) + \epsilon_s S(\mathbf{x},t) \dx P(\mathbf{x},t) \right].
\end{equation}

\section{Derivation of kinetic equations}
\label{sec:app2}
\begin{figure}[!here]
\begin{center}
\includegraphics[width=7cm]{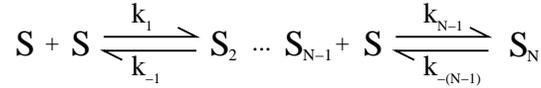}
\caption{Reaction scheme.}
\label{fig:13}
\end{center}
\end{figure}

The reaction scheme is described in Fig \ref{fig:13}. We have 
\begin{equation}
\frac{d[S_N]}{dt} = k_{N-1}[S][S_{N-1}]-k_N[S_N].
\label{eq:app1}
\end{equation}
and
\begin{equation}
\frac{d[S_n]}{dt}=k_{n-1}[S][S_{n-1}]-\left( k_{-(n-1)}+k_n [S] \right)[S_n]+k_{-n}[S_{n+1}],~~n=2,\ldots,(N-1).
\end{equation}
We us assume that all intermediate states are in equilibrium. This assumption implies that intermediate reactions are very fast. With this assumption we get
\begin{equation}
k_{(n-1)}[S][S_{(n-1)}]-k_{-(n-1)}[S_n] =0, n=2,\ldots,N-2,
\end{equation}
\begin{equation}
[S_n] = \frac{k_{n-1}}{k_{-(n-1)}}[S][S_{n-1}]= K_n [S][S_{n-1}],~n=2,3,\ldots,N-2
\label{eq:app2}
\end{equation}
where $K_n=k_n/k_{-n}$.

Using Eq. (\ref{eq:app2}) and Eq. (\ref{eq:app1}) we obtain
\begin{equation}
\frac{d[S_N]}{dt} = k_{clus.}[S]^N-k_N[S_N].
\label{eq:app1}
\end{equation}
In another approximation we replace $[S_N]$ by $\sum^N_{n=2} S_n= P$ to get the equation.
\section{Green's functions}
\label{sec:app3}
The Green's function in Eq. (\ref{eq:9a}) and Eq. (\ref{eq:9b}) has the following form
\begin{equation}
G_\alpha(s) =\left[s-D_\alpha\dx^2+v_\alpha \mbox{sgn}(x) \dx\right]^{-1}
\end{equation}
\begin{eqnarray}
G_\alpha(x|x') = \left\{\begin{array}{ll}
                        \ex(-\rho_{\alpha}(x-x'))A_\alpha(x,x') & \eif x<x'<0 \\
                        \ex(-\rho_{\alpha}(x-x'))A_\alpha(x',x) & \eif x'<x<0, \\ 
                        \frac{1}{2D_\alpha(-\rho_\alpha+\gamma_\alpha)} \ex(\rho_\alpha(x+x')-\gamma_\alpha(x-x')) & \eif x'<0<x,                 
                      \end{array} \right.
\label{eq:grn1}
\end{eqnarray}
 and
\begin{eqnarray}
G_\alpha(x|x') = \left\{\begin{array}{ll}
                        \frac{1}{2D_\alpha(-\rho_\alpha+\gamma_\alpha)} \ex(-\rho_\alpha(x+x')-\gamma_\alpha(x'-x)) & \eif x<0<x',\\
                        \ex(\rho_{\alpha}(x-x'))B_\alpha(x,x') & \eif 0<x<x'\\
                        \ex(\rho_{\alpha}(x-x'))B_\alpha(x',x) & \eif 0<x'<x, \\                                     
                      \end{array} \right.
\label{eq:grn2}
\end{eqnarray}
where  $A_\alpha(x,x')=\frac{1}{2D_\alpha \gamma_\alpha}\left(\ex(\gamma_{\alpha}(x-x'))-\frac{\rho_{\alpha}}{\rho_{\alpha}-\gamma_{\alpha}}\ex(\gamma_{\alpha}(x+x'))\right)$ and  $B_\alpha(x,x')=\frac{1}{2D_\alpha \gamma_\alpha}\left(\ex(\gamma_{\alpha}(x-x'))-\frac{\rho_{\alpha}}{\rho_{\alpha}-\gamma_{\alpha}}\ex(-\gamma_{\alpha}(x+x'))\right)$. Here $\rho_\alpha=v_\alpha/(2D_\alpha)$ and $\gamma_\alpha = \sqrt{\rho_\alpha^2+s/D_\alpha}$ are constants.\\


\begin{thebibliography}{29}
\bibitem{sgar}A Sgarlata, PD Szkutnik, A Balzarotti, N Motta and F Rosei 2003 Appl. Phys. Lett. {\bf 83}, 4002 (2003).
\bibitem{omi}H Omi, T Ogino, Thin Solid Films 369, {\bf 88} (2000).
\bibitem{ogi}T Ogino, H Hibino, Y Homma, Y Kobayashi, K Prabhakaran, K Sumitomo, H Omi, Acc. Chem. Res. {\bf 32}, 447 (1999).
\bibitem{sekar}K Sekar, G Kuiri, P V Satyam, B Sundaravel D P mahapatra and B N Dev, Surf. Sc. {\bf 339}, 96 (1995). 
\bibitem{kim}Kim H J, Zhao Z M and Xie Y H Phys. Rev. B {\bf 68} 205312 (2003).
\bibitem{xie}YH Xie, SB Samavedam, M Bulsara, TA Langdo, EA Fitzgerald, Appl. Phys. Lett. {\bf 71}, 3567 (1997).
\bibitem{kim2}HJ Kim, JY Chang, YH Xie, J. Cryst. Growth {\bf 247}, 251 (2003).
\bibitem{das}AK Das, J Kamila, BN Dev, B Sundaravel, G Kuri, Appl. Phys. Lett. {\bf 77}, 951 (2000).
\bibitem{sch}Th Schmidt, JI Flege, S Gangopadhyay, T Clausen, A. Locatelli, S. Heun, J. Falta, Phys. Rev. Lett. {\bf 98}, 066104 (2007).
\bibitem{sch2}Th Schmidt, S Gangopadhyay, JI Flege, T Clausen, A Locatelli, S Heun, J Falta, New J. Phys. {\bf 7}, 193 (2005).

\bibitem{trilo1} Roy A, Bagarti T, Bhattacharjee K, Kundu K and Dev B N 2012 Surf. Sc. {\bf 606} 777-783.
\bibitem{trilo2} Bagarti T, Roy A, Kundu K and Dev B N 2012 AIP Advances {\bf 2} 042101.

\bibitem{hav}Havlin S and ben-Avraham D 1987 Adv. Phys. {\bf 36}, 695-798.
\bibitem{tait}Taitelbaum H and Koza Z 2000 Physica A {\bf 285} 166-175.
\bibitem{jay}P. K. Datta and A. M. Jayannavar 1992 Pramana-J. Phys. {\bf 38} 257-269.
\bibitem{gras}P. Grassberger and I. Procaccia 1982 J. Chem. Phys. {\bf 77} 6281-6284.
\bibitem{thm} Th. M. Nieuwenhuizen and H. Brand 1990 J. Stat. Phys. {\bf 59} 53-72.
\bibitem{guil}G. Abramson and H. S. Wio 1995 Chaos Soliton and Fract. {\bf 6} 1-5.


\bibitem{mkuntz1} K\"{u}ntz M and Lavall\'{e}e P 2003 J. Phys. D: Appl. Phys. {\bf 36} 1135-1142.
\bibitem{mkuntz2}K\"{u}ntz M and Lavall\'{e}e P 2004 J. Phys. D: Appl. Phys. {\bf 37} L5-L8.

\bibitem{tdfrank1}Frank T D 2002 Phys. Lett. A {\bf 305} 150-159.
\bibitem{tdfrank2}Frank T D 2005 Nonlinear Fokker–Planck Equations Springer Berlin.
\bibitem{lisa1}Borland L 1998 Phys. Rev. E {\bf 57} 6634-6642.
\bibitem{lisa2}Borland L, Pennini F, Plastino A R and Plastino A 1999 Eur. Phys. J. B {\bf 12} 285-297.
\bibitem{mtz}Martinez S, Plastino A R and Plastino A 1998 Physica A {\bf 259} 183-192.

\bibitem{frndo}Fernando A E, K A Landman and M J Simpson 2010 Phys. Rev. E {\bf 81} 011903.
\bibitem{mart}Burger M, Di Francesco M, Jan-Frederik Pietschmann and Schlake B 2010 SIAM J. Math. Anal. {\bf 42} 2842-2871.
\bibitem{kwon}Kwon S and Kim Y 2011 Phys. Rev. E {\bf 84} 041103.
\bibitem{kania}Kaniadakis G and Quarati P 1993 Phys. Rev. E {\bf 48} 4263-4270.

\bibitem{niraj1}Kumar N and Horsthemke W 2011 Phys. Rev. E {\bf 83} 036105.
\bibitem{niraj2}Kumar N and Kenkre V M 2008 PNAS {\bf 105} 18752-57.
\bibitem{eco}Economou E N 2006 Green's Functions in Quantum Physics third ed. Springer Berlin Heidelberg.
\bibitem{hilde}Hildebrand F B Methods of Applied Mathematics second ed. Prentice-Hall New Jersey.
\bibitem{tal}Abate J, Valko P P 2004 Int. J. Numerical Meth. Engng. {\bf 60} 979-993.
\bibitem{len} A length $L$ is introduced which defines the domain of integration which is taken much larger than the diffusion length. Neumann boundary condition at the boundaries will not affect the short time solution.
\bibitem{vanK} van Kampen N G 1981 Stochastic Processes in Physics and Chemistry first ed. North-Holland Amsterdam.
\bibitem{radek} R Erban, S J Chapman, and P Maini 2007 A practical guide to stochastic simulations of reactiondiﬀusion processes 35 pages available as http://arxiv.org/abs/0704.1908.
\bibitem{ames}Ames W F 1977 Numerical Methods for Partial Differential Equations second ed. Academic Press Inc. New York.
\bibitem{chaklub}Chaikin P M, Lubensky T C 1998 Principles of condensed matter physics first ed. Cambridge University Press.
\bibitem{nayf}Nayfeh A H 1973 Perturbation Methods John Wiley \& Sons Inc New York.


\end{thebibliography}
\end{document}